\documentstyle[preprint,aps]{revtex}
\begin{document}

%
%

\title{{\it Ab initio} molecular dynamics via density based energy functionals}
\author{Vaishali Shah, Dinesh Nehete and D. G. Kanhere$^\dagger$}
\address{Department of Physics, University of Poona, Pune 411 007, India}
\maketitle

\begin{abstract}

The use of energy functionals based on density as the basic variable
is advocated for {\it ab initio} molecular dynamics.
It is demonstrated that the constraint of positivity of density can be
incorporated easily by using square root density for minimization of the
energy functional.
An {\it ad hoc} prescription for including nonlocal pseudopotentials
for plane wave based calculations is proposed and is shown to yield
improved results.
Applications are reported for equilibrium geometries of small finite
systems, {\it viz.} dimers and trimers of simple metal atoms like Na and Mg,
which represent a rather stringent test for approximate kinetic energy
functionals involved in such calculations.


\end{abstract}

\newpage

%
%

\section {Introduction}

First principle density functional based molecular dynamics (DF-MD),
initiated by Car and Parrinello (CP) \cite{car,rem}, has become a powerful
technique for {\it ab initio} investigations of a number of properties
of clusters and extended systems \cite{mcp}.
This technique, which unifies the conventional density functional theory (DFT)
with classical molecular dynamics, views the problem as that of total
energy minimization involving electronic and ionic degrees of freedom.
This is achieved via simulated annealing implemented through Langrangian
equations of motion, which are fictituous for electronic degrees of freedom
and real for ionic coordinates.
The method has also given impetus to the development of better and faster
techniques for large scale electronic structure calculations with fixed
geometry, {\it e.g.} clusters involving large number of atoms \cite{sti}.
The applications based on the CP formalism fall into two broad categories:
\begin{enumerate}
\item
{\it ab initio} prediction of ground state properties like equilibrium
geometry, and
\item
finite temperature properties obtained via trajectories of the
system moving on Born-Oppenheimer (BO) surface.
\end{enumerate}
A majority of the applications belong to the first category.
Inspite of a number of technical advances like accelerated algorithms
involving real space approach \cite{ksm}, preconditioned conjugate
gradient minimization methods \cite{mtp}, and analytically continued energy
functionals \cite{tar}, it is clear that the method becomes prohibitively
expensive for large scale calculations.
Typically these algorithms scale as $O(N_o^2N_b)$ where $N_o$ is the number of
orbitals and $N_b$ is the number of basis functions, $N_o^2N_b$ being
the dominant cost of orthogonalization of electronic orbitals.
In addition, for simple metal systems, a tricky problem of charge sloshing
\cite{mtp} has also been noticed which limits the timestep that can be used
in dynamics.
Recently, there have been a few attempts towards obtaining
linear scaling, {\it e.g.} a non-orthogonal localized basis formulation
\cite{gal},
finite difference real space discretization coupled with recursion
method \cite{bar} and density matrix based formulation.

In the present work, following the Hohenberg-Kohn theorem \cite{hoh,ksh},
we advocate a rather simple method in which the total energy functional
is written in terms of density as the basic variable.
Such an orbital free method (OFM) has been used by Pearson {\it et al}
\cite{pear}.
They have used this method for calculating equilibrium lattice
separation, bulk modulus and vacancy formation energy of solid Sodium.
The method has also been applied to the calculation of free energies
of vacancies \cite{mad}.
As pointed out in their work, at least for simple metallic systems the method
scales linearly with the number of atoms $N_a$ and is capable of treating large
simulation timesteps.
The accuracy of this method hinges upon the correct description
of the kinetic energy functional.
Usually the kinetic energy functional is taken to be the Thomas-Fermi
type with the appropriately scaled Weizsacker correction.
This can be further improved by taking into account linear response properties,
leading to the Perrot form \cite{mad}.
However, there are a few points which must be critically examined.
Firstly, the usual DF-MD methods are implemented using first principles
pseudopotentials which are necessarily nonlocal.
So far, the available orbital free formulation \cite{pear,mad} uses local
potentials only.
Secondly, during the minimization process the positivity of the charge
density $\rho({\bf r})$ must be strictly maintained.
Thirdly, the accuracy of the results, at least so far as the bondlengths are
concerned, which are crucial for ground state geometries and other structural
properties, should be thoroughly assessed.

The present work is motivated by a desire to investigate these questions.
In this work,
\begin{enumerate}
\item
We propose an {\it ad hoc} prescription for incorporating the nonlocal
contribution of the pseudopotential, which is missing from the earlier work.
\item
We use square root density $\sqrt{\rho({\bf r})}$ as the basic variable to
incorporate the positivity constraint on density.
\item
We present results of applications of this method to the equilibrium
bondlengths of dimer and trimer systems. This would be a stringent
test as compared to the applications to extended systems.
\end{enumerate}
We believe that a combination of density based orbital free MD and
Kohn-Sham (KS) orbital based MD may yield a cost effective way for
performing geometry optimization for large clusters.
This can be achieved by first obtaining approximate low temperature
structures by the present method which is $O(N_a)$ and then performing full
MD or a fast quench.
As the kinetic energy functional improves, this way of geometry optimization
should turn out to be a computationally tractable alternative for large
clusters consisting of more than few hundred atoms.
It may also be used for investigating the thermodynamic properties of
large scale systems where the conventional methods may be prohibitive
in terms of computer time.
The method will be most useful for simple metal systems provided it yields
acceptable bondlengths.
It is hoped that this kind of work will give impetus to formulating
better kinetic energy functionals.

In the next section we give our formalism and computational details.
In section III we present the results for Na$_2$, Mg$_2$ and Mg$_3$ and
compare them with full self consistent KS calculations.

\section {Formalism and Computational details}

\subsection { Total Energy Calculation}

The total energy of a system of $N_e$ interacting electrons and $N_a$ atoms,
according to the Hohenberg-Kohn theorem, can be uniquely expressed as a
functional of the electron density $\rho({\bf r})$ under an external field
due to the nuclear charges at coordinates ${\bf R}_n$.
\begin{equation}
   E\Bigl[\rho,\{{\bf R}_n\}\Bigr] = T[\rho]
                                  + E_{xc}[\rho]
                                  + E_c[\rho]
                                  + E_{ext}\Bigl[\rho,\{{\bf R}_n\}\Bigr]
                                  + E_{ii}\Bigl(\{{\bf R}_n\}\Bigr)
\end{equation}
where
\begin{equation}
   T[\rho] =   \frac{3}{10}(3\pi^2)^{\frac{2}{3}}
               \int{\rho({\bf r})^{5/3} d^3r}
            +  \frac{\lambda}{8} \int{\frac{\nabla\rho({\bf r})
               \cdot\nabla\rho({\bf r}) d^3r}{\rho({\bf r})}}
\end{equation}
is the kinetic energy functional.
The first term in this functional is the Thomas-Fermi term, exact in
the limit of homogenous density, and the second is the gradient correction
due to Weizsacker.
It has been pointed out that instead of $\lambda = 1$ which
is the original Weizsacker value, $\lambda = \frac{1}{9}$ and other
empirical values turn out to yield better results \cite{par}.
The exchange-correlation energy in the usual local density approximation
is given by
\begin{equation}
   E_{xc}[\rho] =   \int{\rho({\bf r})
                                       \varepsilon_{xc}(\rho) d^3r}
\end{equation}
where $\varepsilon_{xc}[\rho]$ denotes the exchange and correlation energy
per particle of a uniform electron gas of density $\rho$.
\begin{equation}
   E_c[\rho] = \frac {1}{2}\int\int{\frac{\rho({\bf r}) \rho({\bf r}^\prime)}
                     {\vert{{\bf r}-{\bf r}^{\prime}}\vert}}d^3rd^3r^\prime
\end{equation}
is the electron-electron Coulomb interaction and
\begin{equation}
   E_{ext}\Bigl[\rho,\{{\bf R}_n\}\Bigr] =  \int{V({\bf r})
                                                 \rho({\bf r}) d^3r}
\end{equation}
is the electron-ion interaction where $V({\bf r})$ is the external potential
and the last term in the Eq. (1), $E_{ii}$, denotes the ion-ion
interaction energy.

The external potential is usually taken to be a convenient pseudopotential,
and in general, it is nonlocal. Let us recall that in the standard
pseudopotential formulation \cite{ihm} the nonlocal contribution to the
electron-ion energy for an electron in the state $\Psi({\bf r})$ is given by
\begin{equation}
    E_{nl}[\Psi] = \int{\Psi({\bf r}) V_{ps,l}({\bf r})\widehat{P_l}
                                      \Psi({\bf r}) d^3r}
\end{equation}
where $V_{ps,l}({\bf r})$ is the $l$-dependent part of the pseudopotential
and $\widehat{P_l}$ is the angular momentum projection operator.
In analogy, we suggest the following expression for nonlocal contribution
to the total energy in terms of square root density.
Let
\begin{equation}
   E_{nl}[\tilde{\rho}] = \int{\tilde{\rho}({\bf r})
                          V_{ps,l} ({\bf r}) \widehat{P_l}
                          \tilde{\rho}({\bf r}) d^3r}
\end{equation}
where
\begin{equation}
     \tilde{\rho}({\bf r}) = \sqrt{\rho({\bf r})}.
\end{equation}
The exchange-correlation energy is calculated using the Ceperley-Alder
exchange potential as interpolated by Perdew-Zunger \cite{per}.
Although in the present work we use a simple kinetic energy
(KE) functional, improved KE functionals useful at least for simple
metals have been reported.
For example, Smargiassi and Madden \cite{mad} have investigated a
family of KE functionals, with applications to Na and Al, giving
comparable accuracy as obtained by KS method.

\subsection {Dynamics}

Typically the MD procedure proceeds via two steps.
The first step is to obtain the ground state energy $E$ and density
$\rho({\bf r})$ for a fixed geometry, which could be done via CP
dynamics performed on electronic degrees of freedom only.
However, gradient based minimization techniques have also been found
to be effective in locating the minimum of general functions
and are known to have fast convergence \cite{tet}.
We have applied the conjugate gradient (CG) algorithm \cite{num,gill,psh} for
minimizing the total energy functional for a fixed geometry configuration.
Thus, starting with a trial {\bf r}-space charge density $\rho^0$, the CG
algorithm proceeds as follows:
the new charge density for the $k+1^{th}$ iteration is constructed
by linearly combining the charge density $\rho^k({\bf r})$ and direction
$d^k({\bf r})$ for the $k^{th}$ iteration as
\begin{equation}
    \rho^{k+1}({\bf r}) = \rho^k({\bf r}) + \alpha^kd^k({\bf r})
\end{equation}
The initial search direction is taken to be the steepest descent direction
\begin{equation}
   d^0({\bf r}) = - g^0({\bf r}),
\end{equation}
and successive conjugate directions $d^k({\bf r})$ are defined as
\begin{equation}
    d^k({\bf r}) = -g^k({\bf r}) + \beta^k d^{k-1}({\bf r}),
\end{equation}
where
\begin{equation}
    g^k({\bf r}) = \frac{\delta{E[\rho^k]}}{\delta{\rho^k({\bf r})}}
\end{equation}
and
\begin{equation}
    \beta^k = \frac{g^k{\cdot}g^k}{g^{k-1}{\cdot}g^{k-1}}.
\end{equation}
The search parameter $\alpha^k > 0$ is chosen to minimize the functional
$E^{\prime}(\alpha) = E[\rho^k + \alpha d^k]$ for a given $\rho^k$
and $d^k$.
In the present case, we perform this line minimization numerically,
using Brent's method \cite{num}, and we notice that it does not
turn out to be very expensive (about 7 function evaluations to locate
an $\alpha^k$).
It is well known that subsequent minimizations along the CG
directions tend to introduce errors in the calculation due to finite
precision.
We have found it advantageous to restart the CG search after every
few iterations with a steepest descent direction.
To incorporate the positivity constraint on density into the minimization
procedure, we vary $\tilde{\rho}({\bf r})$ (Eq. (8)) rather than
$\rho({\bf r})$, with Eq. (9)-(12) reexpressed in terms of
$\tilde{\rho}({\bf r})$.

After performing minimization to a desired degree of convergence,
trajectories of ions and fictitious electron dynamics are simulated
using Langrage's equations of motion with Verlet algorithm.
To simulate the motion on the BO surface, we start with the
Lagrangian defined by
\begin{equation}
      L = K_e + K_a - E\Bigl[\tilde{\rho},\{{\bf R}_n\}\Bigr]
\end{equation}
where
\begin{equation}
      K_e = \mu \int{\dot{\tilde{\rho}}({\bf r}) d^3r}
\end{equation}
and
\begin{equation}
      K_a = \frac {1}{2} \sum_n {M_n \vert {\dot{R}_n(t)}\vert^2}
\end{equation}
is the kinetic energy of electrons and the kinetic energy of the ions
placed at $\{{\bf R}_n\}$ respectively.
The dot denotes time derivative and $\mu$ and $M_n$ are respectively the
fictitious mass of electrons and the mass of the $n$-th atom.
The fictitious mass of the electrons $\mu$ is a parameter to model
the classical motion of density analogous to atomic motion.
The above Lagrangian leads to the following equations of motion
\begin{equation}
\mu \ddot{\tilde{\rho}}({\bf r},t) = - \frac {\delta E}{\delta \tilde{\rho}
                                                           ({\bf r},t)}
\end{equation}
\begin{equation}
M_n \ddot{{\bf R}}_n(t) = - \nabla_n E
\end{equation}
for electrons and ions respectively, subject to the constraint
\begin{equation}
      N_e =  \int{\tilde{\rho}^2({\bf r}) d^3r}.
\end{equation}

The dynamics being conservative the grand total energy $E_{GT}$ is a
conserved quantity of motion.
Thus,
\begin{equation}
 E_{GT} = E_e + E_a + E\Bigl[\tilde {\rho},\{{\bf R}_n\}\Bigr]
\end{equation}
is the sum of the fictitious kinetic energy of electrons, kinetic
energy of ions and total energy of electrons.
The grand total energy is the parameter used to monitor and judge the
quality of DF-MD numerical simulations.
The greatest advantage of using density as the variational parameter is that
the constraint of orthonogonality of wavefunctions can be done away with.
This saves considerable amount of computation, thus making the
dynamics fast.
An alternative to CP algorithm for dynamical simulations of ionic systems
has been suggested by Payne {\it et al} \cite{mtp}. In this method the
electronic degrees of freedom are relaxed to the instantaneous ground state
at the new ionic coordinates.
We have used this CG-MD procedure along with the density predictor method
\cite{mcp} for dynamical simulations.
The density in successive timesteps is constructed using the first order
density predictor as
\begin{equation}
    \tilde{\rho}\prime\Bigl(\{{\bf R}_n(t+1)\}\Bigr) =
                                \tilde{\rho}\Bigl(\{{\bf R}_n(t)\}\Bigr)
                              + \Bigl[\tilde{\rho}\Bigl(\{{\bf R}_n(t)\}\Bigr)
                              - \tilde{\rho}\Bigl(\{{\bf
R}_n(t-1)\}\Bigr)\Bigr]
\end{equation}
where $\tilde{\rho}\prime$ denotes a trial density which is then used for
further minimization.
This reduces the number of CG minimization steps by a factor of 2
(typically 2 or 3 CG steps have been found to be sufficient for convergence).

The calculations for  Na$_2$, Mg$_2$ and Mg$_3$ have been performed on a
periodically repeated unit cell of length 35 a.u.
with a 48 $\times$ 48 $\times$ 48 mesh.
The square root charge density is expanded in terms of plane waves as
\begin{equation}
     \tilde{\rho}({\bf r}) = \sum_{{\bf G}}\tilde{\rho}({\bf G})e^{i{\bf G}
                                                    {\cdot}{\bf r}}.
\end{equation}
CG dynamics involves the calculation
of the first derivatives of energies with respect to $\tilde{\rho}({\bf r})$.
The electrostatic energy, gradient correction to the kinetic energy and the
nonlocal energy and their respective derivatives were calculated in
Fourier space and then transformed to ${\bf r}$-space.
The MD calculations were performed in the conventional CP technique and
conjugate gradient CP technique.
Local calculations were conveniently performed in the ${\bf r}$-space.
The energy cutoff used for local calculations was 30 Rydbergs and that
for nonlocal energy calculations 10 Rydbergs.
The dimer dynamics with conjugate gradient CP technique is quite stable
with one CG step after adjusting the density by the predictor
method for timestep of 10 a.u. and $\mu$ = 600 a.u.
Calculations with a timestep of 50 a.u. required 4 CG steps for the
trajectories to remain on the BO surface.

\section {Results and Discussion}

In this section, we present our results for the bondlengths of
 Na$_2$, Mg$_2$, Mg$_3$
and compare them with that obtained via conventional {\it
ab initio} MD. We have performed both fixed geometry minimization as well
as simulated annealing MD. All the results presented here are obtained
with energy convergence upto $10^{-13}$ for total energy minimization

Table I shows the comparison of bondlengths via the OFM
with local and nonlocal pseudopotentials and KS self consistent formulation.
It is gratifying to note that the maximum error in the bondlength with
nonlocal pseudopotential is of the order of 2\%.
It can be seen that nonlocality improves the bondlengths
considerably. However, this is at the cost of additional operations which
go as $N_a \times N_b$.

In order to understand the total energy behaviour using OFM and also
the effect of nonlocality, we have plotted the total energy as a
function of interatomic separartion in Fig. 1. The curve labeled A
corresponds to the KS local results, B corresponds to the KS nonlocal,
C and D represent the corresponding results obtained by the present method.
The energy and the distances are measured with respect to the equilibrium
quantities obtained by respective calculations. It can be seen that although
the magnitude of changes is rather small, the effect of nonlocality is
somewhat drastic in OFM. This leads to a faster approach to
equilibrium with the inclusion of nonlocality. However, it also shows a
worrisome feature, namely, the effect of nonlocality in the OFM is
considerably more as compared to the full KS calculation.
To assess the quality of the charge density obtained via the OFM, we have
plotted it in Fig. 2 along with that due to the KS method. As can be  seen,
there is an overall agreement in the nature of the density curve. The OF
density is overestimated in the region away from the atomic site and
underestimated at the atomic site. This is understandably due to
the inexact formulation of kinetic energy functionals. However,
the long range behaviour is identical in both the cases.
The variation of total energy $E$ (continuous line) and $E_{GT}$
(dashed line) with time during the free MD simulation run for Mg$_2$
is shown in Fig. 3.
It can be seen that the grand total energy $E_{GT}$ is constant
throughout the simulation to an order of 10$^{-6}$. These results have been
obtained  by CG-MD using density predictor method. We have observed that
for the dimer case a much higher timestep $\sim$ 50a.u. also keeps the
system on Born-Oppenheimer surface with a slight increase in the
computational cost. It may be mentioned that the Langragian dynamics
(CP Technique) is much faster and our results for a timestep
$\sim$ 20 are more or less identical with the CG dynamics results.

\section {Conclusion}

In this work we have presented a fast but approximate density based {\it ab
initio} MD and demonstrated that the bondlengths for dimers and trimers are
obtained to within an accuracy of 2\% for nonlocal and $\sim$ 10\% for
local calculations. We have shown that the CG technique in conjugation with
a simple density predictor method allows us to use large timesteps
for dynamics. The fixed geometry minimization using CG can be obtained to
a high degree of accuracy within less than 200 iterations.
Yet another alternative to incorporate the positivity constraint is by
constraining variation of density during the minimization (Eq. (9)) by
restricting $\alpha^k$ to appropriate range \cite{psh}.
Our preliminary investigation indicates that this technique is much faster
compared to the square root density minimization. This is understandable,
because it is well known that CG works best for quadratic and near-quadratic
functionals, and the degree of nonlinearity of the energy functional is
reduced if expressed in terms of $\rho$ rather than $\tilde{\rho}$.

We believe the present results on small systems to be a stringent test
and the method to be a viable alternative for calculating finite temperature
properties of systems involving a large number of atoms.

\section {Acknowledgements}
It is a pleasure to acknowledge Mihir Arjunwadkar for a number of discussions
and help in the initial stages of the work. Partial financial assistance from
Department of Science and Technology (DST), Government of India and Centre for
Development of Advanced Computing (C-DAC), Pune is gratefully acknowledged.
Two of us (V. S and D. N) acknowledge financial assistance from C-DAC.

\newpage

\vspace{3pt}

TABLE I. Equilibrium bonglength obtained via OFM local
formulation (column 2) and OFM nonlocal formualtion (column 3) compared with
that of KS self consistent method (column 4)
(all values in atomic units).

\vspace{0.5in}
\begin{tabular}{lllr} \hline\hline
{}~~~~~~~~  System   & ~~      OFM local   &~~      OFM nonlocal &~~~~
KS nonlocal   \\  \hline
{}~~~~~~~~~~ Na$_2$  & ~~~~~~    5.34      &~~~~~~~        5.62  &  5.68 ~~~~~~
           \\
{}~~~~~~~~~~ Mg$_2$  & ~~~~~~    5.35      &~~~~~~~        6.32  &  6.26 ~~~~~~
           \\
{}~~~~~~~~~~ Mg$_3$  & ~~~~~~    5.44      &~~~~~~~        5.88  &  5.82 ~~~~~~
           \\  \hline\hline
\end{tabular}

\newpage
\centerline{\bf FIGURES}

\vspace{1.0in}
FIG. 1. The behaviour of total energy near equilibrium separation
as a function of distance. All distances and energies are measured with
respect to their corresponding equilibrium values.
Curve A:  KS local, B:  KS nonlocal, C:  OFM local, D:  OFM nonlocal.

\vspace{1.0in}
FIG. 2. charge density $r^2 \rho({\bf r})$ along the symmetry axis for
Mg$_2$. The dashed line corresponds to the KS result and the continuous
line represents the OFM result.

\vspace{1.0in}
FIG. 3. Time evolution of the total energy $E$ (solid line) and grand
total energy $E_{GT}$ (dashed line) of Mg$_2$ during a 1000
step free OFM MD run with timestep of 10 a.u.


\begin{references}

\bibitem[\dagger]{email}
Electronic address: kanhere@parcom.ernet.in

\bibitem{car}
R. Car, M. Parrinello, Phys. Rev. Lett., {\bf 55}, 685(1985)

\bibitem{rem}
D. K. Remler and P. A. Madden, Molecular Physics, {\bf 70}, 921(1990)

\bibitem{mcp}
T. A. Arias, M. C. Payne and J. D. Joannopoulos, Phys. Rev. B.
{\bf 45}, 1538(1992)

\bibitem{sti}
I. Stitch, R. Car, M. Parrinello and S. Baroni, Phys. Rev. B
 {\bf 39}, 4997(1989)

\bibitem{ksm}
R. D. King-Smith, M. C. Payne and J. S. Lin, Phys. Rev B
{\bf 44}, 13063(1991)

\bibitem{mtp}
M. Payne, M. P. Teter, D. C. Allan, T. A. Arias and J. D. Joannopoulos,
Rev. Mod. Phys., {\bf 64}, 1045(1992)

\bibitem{tar}
T. A. Arias, M. C. Payne and J. D. Joannopoulos, Phys. Rev. Lett.
{\bf 69}, 1007(1992)

\bibitem{gal}
G. Galli and M. Parrinello, Phys. Rev. Lett. { \bf 69}, 3547(1992)

\bibitem{bar}
S. Baroni and P. Giannozzi, Europhys. Lett. {\bf 17}, 547(1992)

\bibitem{hoh}
P. Hohenberg and W. Kohn, Phys. Rev. {\bf 136}, B864(1964)

\bibitem{ksh}
W. Kohn and L. J. Sham, Phys. Rev. {\bf 140}, A1133(1965)

\bibitem{pear}
M. Pearson, E. Smargiassi and P.A. Madden, J. Phys. Condens.
Matter {\bf 5} 3221 (1993)

\bibitem{mad}
E. Smargiassi and P. A. Madden, Phys. Rev. B. {\bf 49}, 5220(1994)

\bibitem{par}
R. G. Parr and W. Yang, {\it Density Functional Theory of Atoms
and Molecules} (O. U. P., Oxford, 1989)

\bibitem{ihm}
J. Ihm, A. Zunger and M. L. Cohen, J. Phys. C:Solid State Phys.,
{\bf 12}, 4409(1979)

\bibitem{per}
J. P. Perdew and A. Zunger, Phys. Rev. B{\bf 23}, 5048(1981);
U von Barth and L. Hedin, J. Phys. C: Solid State Phys.
{\bf 5}, 1629(1972)

\bibitem{tet}
M. P. Teter, M. C. Payne and D. C. Allan, Phys. Rev. B{\bf 40}, 12255(1989)

\bibitem{num}
W. H. Press, B. P. Flannery, S. A. Teukolsky, W. T. Vetterling,
{\it Numerical Recipes}, Cambridge University Press, Cambridge (1987)

\bibitem{gill}
P. E. Gill, W. Murray, M. H. Wright, {\it Practical Optimization},
Acamedic Press, London (1988)

\bibitem{psh}
B. N. Pshenichny and Y. M. Danilin, {\it Numerical Methods in Extremal
Problems}, Mir publishers, Moscow (1978)

\end{references}
\end{document}